\documentclass[prb,twocolumn]{revtex4-1}
\usepackage{amssymb,amsmath}
\usepackage{graphicx}
\def\s{\sigma}
\def\ra{\rangle}
\def\la{\langle}

\newcommand{\eq}[1]{Eq.~(\ref{#1})}
\newcommand{\fig}[1]{Fig.~\ref{#1}}
\newcommand{\op}[1]{\hat{#1}}
\newcommand{\opd}[1]{\hat{#1}^\dag}

\begin{document}

\title{Signatures of spin blockade in the optical response of a charged quantum dot}
\author{E. G. Kavousanaki}
\author{Guido Burkard}
\affiliation{Department of Physics, University of Konstanz, D-78464 Konstanz, Germany}
\date{\today}

\begin{abstract}
We model spin blockade for optically excited electrons and holes
in a charged semiconductor quantum dot.
We study the case where the quantum dot is initially charged with a
single electron and is then filled with an additional, optically excited
electron-hole pair, thus forming a charged exciton (trion).
To make contact with recent experiments, we model an optical pump-probe setup,
in which the two lowest quantum dot levels (s and p shells) are
photoexcited.
Using the Lindblad master equation,
we calculate the differential transmission 
spectrum as a function of the pump-probe time delay.
Taking into account both spin conserving and spin-flip 
intraband relaxation processes, we find that the presence of the
ground-state electron spin leads to an optical spin blockade at short delay
times which is visible as a crossover between two exponential decays
of the differential transmission. 
To make predictions for future experiments, we also study the
dependence of the spin-blockade on an external magnetic field. 
\end{abstract}

\maketitle


\section{Introduction}

One of the promising solid-state implementations for the realization of
quantum computing that has been under intense study over the past years involves 
the use of a single electron spin confined to a charged quantum dot (QD)\cite{Loss1998}.
The discrete QD energy structure allows for long spin lifetimes, e.g. exceeding
one second in electrically defined GaAs QDs \cite{Hanson2007}, in comparison with the bulk materials 
or semiconductor nanostructures of higher dimension.
However, it has been shown that the inhomogeneous dephasing time $T_2^\ast$ in
GaAs QDs in the presence of an unpolarized ensemble of nuclear spins
in the QD material is of the order of $\sim10$ ns, while the intrinsic spin
coherence time $T_2$ can reach values beyond 1~$\mu$s \cite{Petta2005a}. 
The decoherence time is relevant for quantum information applications 
where it should exceed the elementary quantum gate operation time by
a substantial factor.
Electrical control of single spins has been realized in
timescales of about 50 to 100 nanoseconds \cite{Koppens2006,Nowack2007},
while
ultrafast optical pulses have been shown to allow ensemble spin manipulation in 
picosecond timescales \cite{Greilich2006,Greilich2007} and arbitrary coherent
single-spin rotations \cite{Berezovsky2008}.

\begin{figure}
\includegraphics[width=\linewidth]{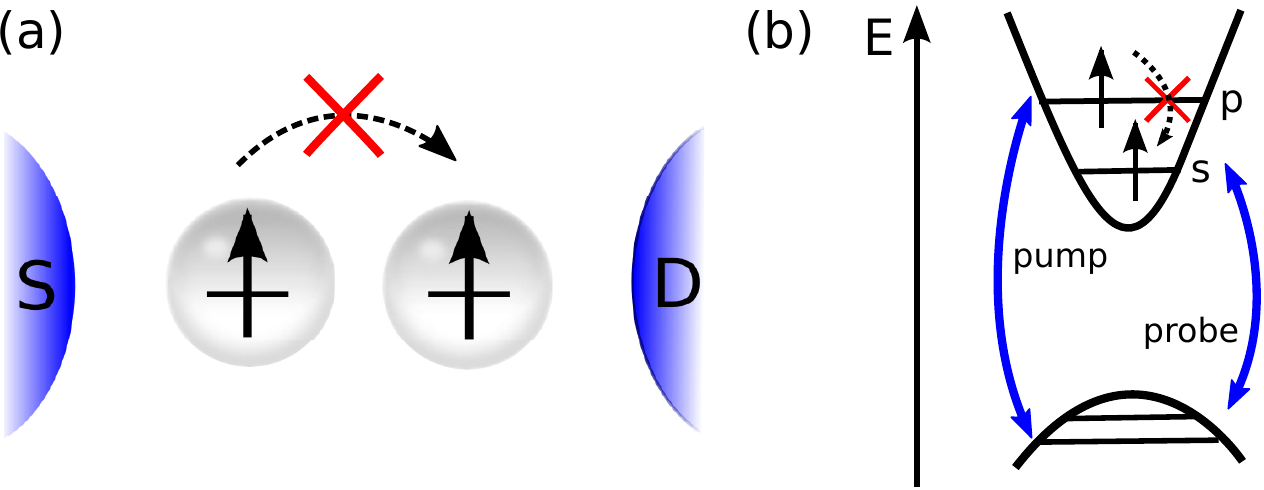}
\caption{
(a) Spin blockade in charge transport through a double QD, connected
to source (S) and drain (D) leads.  A parallel spin configuration (spin
triplet) can lead to a blocking of the current.
(b) The optical analogy of spin blockade in the intraband relaxation
between two QD levels s and p in a pump-probe set-up.
}
\label{SpinBlockade}
\end{figure}

Spin blockade, more generally known as Pauli blockade,
describes a situation where an electronic process is
inhibited for certain spin configurations because the energetically
accessible final states are forbidden by the Pauli exclusion principle
(Fig.~\ref{SpinBlockade}).
In the electric transport between coupled quantum dots, spin blockade
can prevent an electron to
access an energetically favorable path due to spin conservation \cite{Ono2002}
(\fig{SpinBlockade}a).
Observations of the leakage current in the spin-blocking regime
have allowed the study 
of spin decoherence mechanisms and in particular the role of
nuclear spins \cite{Koppens2005,Jouravlev2006}.

In optical experiments, Pauli blocking effects, also
known as phase space filling \cite{HaugKoch2009}, 
are commonly observed in absorption spectra at high 
photoexcitation intensities or when ground state carriers are present.
Spin blockade of the lower Zeeman branch in a singly charged QD
in strong magnetic fields has been studied \cite{Hogele2005},
and in a recent pump-probe experiment \cite{Sotier2009}, signatures
of optical spin blockade have been observed in the
transmission spectra of a charged QD.
Lifting of spin blockade 
is typically more difficult to observe for optically excited carriers
than for transport setups because electron-hole recombination 
processes can be much faster than the spin coherence and relaxation
times.

In this paper, we describe an optical spin blockade effect
in a charged quantum dot with two photoexcited energy levels 
that play the role of the two coupled QDs in transport. 
(Fig.~\ref{SpinBlockade}b)
We show the signature of interlevel (intraband) 
spin relaxation on the differential transmission signal in a 
pump-probe setup (Fig.~\ref{diagram}) and draw
the analogies between optical and transport experiments.
For small QDs, the admixture mechanisms due to spin-orbit coupling
play a smaller role\cite{Khaetskii00PRB} and direct spin-phonon
mechanisms need to be taken into account.

\section{Theoretical Model}

We study a quantum dot in a cubic semiconductor (e.g. GaAs)
charged with a single electron. For self-assembled quantum dots,
lateral dimensions are significantly larger than their height, and
we thus we assume a circular quantum dot in a parabolic confinement
potential characterized by a frequency $\omega$. In analogy with atoms,
single particle eigenstates in QDs are typically labeled as $s$, $p$, $d$,... 
shell, which
for our model correspond to $n=0,1,2,...$ harmonic oscillator states, 
with $n=n_x+n_y$ the total quantum number. Including spin,
single particle states in the conduction band are degenerate with respect
to spin $J_z=\pm 1/2$ in the absence of a magnetic field (for circular QDs).
In the valence band, heavy hole (total angular momentum $J_z=\pm 3/2$) and light 
hole ($J_z=\pm 1/2$) states are split due to confinement by an energy $\Delta_{lh}$. 
Here, we will consider only heavy hole states, assuming that the
heavy hole-light hole mixing near the band edges can be neglected.

\begin{figure}
\includegraphics[width=\linewidth]{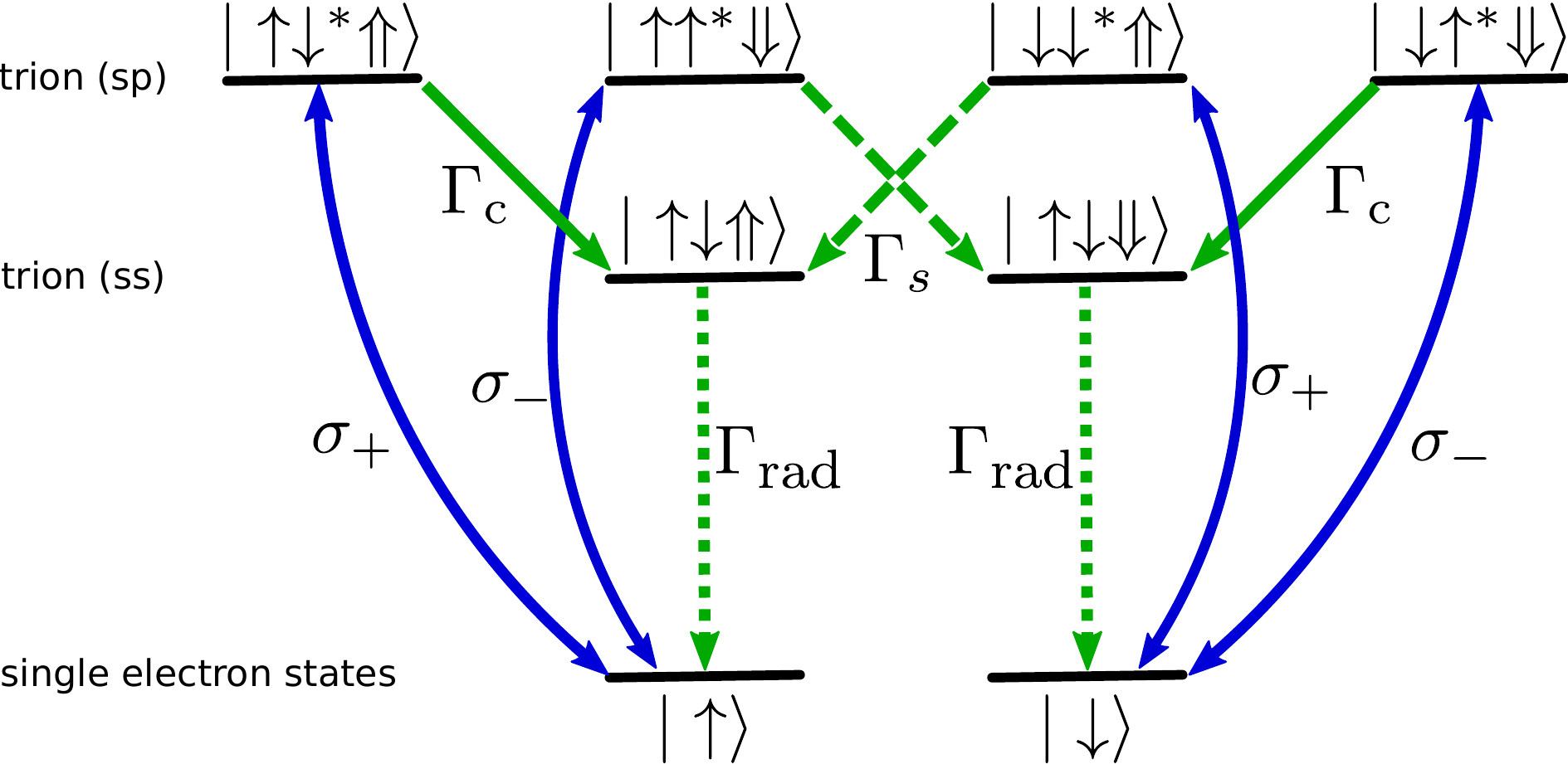}
\caption{
Energy levels of a charged quantum dot under resonant photoexcitation 
with right- (left-) circularly ($\sigma_\pm$) polarized light 
of the first excited level (p shell), indicated by curved blue arrows.
The electron (hole) spin in the lowest QD level (s shell) is denoted by
$\uparrow,\downarrow$ ($\Uparrow,\Downarrow$), while
$\uparrow^*,\downarrow^*$ denotes an electron spin in the excited QD level 
(p shell).
Straight green arrows indicate relaxation processes with $\Gamma_{\rm c}$ 
the intraband spin-conserving relaxation rate,
$\Gamma_{\rm s}$ the intraband spin-flipping rate, and
$\Gamma_{\rm rad}$ the interband radiative recombination rate.
}
\label{diagram}
\end{figure}

The system is optically excited by a strong pump pulse that is resonant to the
first excited QD state ($p$-shell), creating an electron-hole ($e$-$h$) pair with specific
angular momentum depending on the pulse polarization. According to the optical
selection rules, a right- (left-) circularly polarized $\sigma_\pm$ pulse excites a 
$J_z=\mp 1/2$ electron and a $J_z=\pm 3/2$ hole, creating an excited trion state (\fig{diagram}). 
Depending on the spin polarization of the electrons, the (sp) trion state can be an
electron singlet (total trion angular momentum $J_z=\pm 3/2$) 
or triplet ($J_z=\pm 5/2,\pm 3/2, \pm 1/2$)\cite{Bracker2008}. The singlet and triplet states
are split by an energy $\Delta_{ee}$ due to electron-electron exchange interactions, 
which is typically of the order of a few meV. 
In our model, we assume that the pump pulse width is much broader than the singlet-triplet
splitting $\Delta_{ee}$ and thus the latter can be ignored.

We focus on th interlevel relaxation of the photoexcited electron, i.e. relaxation from
the (sp) trion state to the (ss) trion (see \fig{diagram}). Since the latter can only 
be an electron singlet,
the relaxation rate depends strongly on the excited trion state. If it is a spin singlet,
interlevel relaxation takes place through phonon emission on a timescale of a few tens of
ns. On the other hand, if it is a spin triplet,
a spin flip mechanism is required for the relaxation to take place.
This will typically involve spin-orbit coupling in combination with 
phonon emission and take a much longer time as compared to the
spin-conserving relaxation.

In our model, we use the Hamiltonian \cite{HaugKoch2009}
\begin{equation}
H=H_0+H_L+H_C,
\label{Htot}
\end{equation}
where
\begin{equation}
H_0=\sum_{n\s}E_{n\s}^e\opd{e}_{n\s}\op{e}_{n\s}
+\sum_{n\s}E_{n\s}^h\opd{h}_{n\s}\op{h}_{n\s} ,
\end{equation}
describes non-interacting electrons and holes and
\begin{equation}
H_L=-\sum_{n\s}d E(t)\opd{e}_{n\s}\opd{h}_{n\bar{\s}}
-\sum_{n\s}d^\ast E^\ast(t)\op{h}_{n\bar{\s}}\op{e}_{n\s} ,
\end{equation}
is the coupling to the optical field, where
$\opd{e}_{n\s}$ ($\opd{h}_{n\s}$), $\op{e}_{n\s}$ ($\op{h}_{n\s}$) are
the creation and annihilation operators of an electron (hole) in the $n$-th
quantum dot level ($n=s,p$) with spin $\s=\pm\frac{1}{2}$ ($\s=\pm\frac{3}{2}$),
$E_{n\s}^e$ ($E_{n\s}^h$) the single-particle energies of the QD levels,
$d$ the interband dipole moment, and $E(t)$ the electric field.
For a more compact notation, we use the notation 
$\bar{\s}=\uparrow,\downarrow$ when $\s=\downarrow,\uparrow$.

The last term in the Hamiltonian \eq{Htot} describes Coulomb interactions,
\begin{eqnarray}
H_C &=& 
\frac{1}{2}\sum_{nm\s\s'}
V^{ee}_{nm} 
\opd{e}_{n\s}\opd{e}_{m\s'}\op{e}_{m\s'}\op{e}_{n\s}\nonumber\\
&&+\frac{1}{2}\sum_{nm\s\s'} 
V^{hh}_{nm} 
\opd{h}_{n\s}\opd{h}_{m\s'}\op{h}_{m\s'}\op{h}_{n\s}\nonumber\\
&&-\sum_{nm\s\s'} 
V^{eh}_{nm} 
\opd{e}_{n\s}\opd{h}_{m\s'}\op{h}_{m\s'}\op{e}_{n\s}
\end{eqnarray}
where only terms that conserve the number of particles in each QD level are included. 
This is a reasonable approximation for very small QDs in which interlevel spacing 
is much larger than Coulomb interaction. Such terms lead to density
dependent energy shifts, as we will discuss in the next section.

Intraband relaxation of electrons from the $p$ to the $s$ shell
is described with the Lindblad operators
\begin{equation}
L^e_{\s_1\s_2}=\opd{e}_{s\s_1}\op{e}_{p\s_2}.
\end{equation}
Similarly, hole relaxation is described by the operator 
\begin{equation}
L^h_{\s_1\s_2}=\opd{h}_{s\s_1}\op{h}_{p\s_2},
\end{equation}
The dynamics of the density matrix $\rho$ describing the electronic
state of the quantum dot is given by a master equation in the Lindblad form
($\hbar=1$ throughout the paper),
\begin{eqnarray}
\dot\rho
&=&
-i[H,\rho]+
\sum_{\s_1\s_2 r}\Gamma^{r}_{\s_1\s_2}
\left[
L^r_{\s_1\s_2} \rho L^{r\dag}_{\s_1\s_2}
\right.\nonumber\\
&&\left.
-\frac{1}{2} L^{r\dag}_{\s_1\s_2} L^r_{\s_1\s_2} \rho
-\frac{1}{2} \rho L^{r\dag}_{\s_1\s_2} L^r_{\s_1\s_2}
\label{ME}
\right],
\end{eqnarray}
where $r=e,h$ and
\begin{equation}
\Gamma^e_{\s_1\s_2}=
\left\{
\begin{array}{ll}
\Gamma_{\rm c} & \textrm{if}\: \s_1=\s_2\\
\Gamma_{\rm s} & \textrm{if}\: \s_1\ne \s_2\\
\end{array}
\right. 
\end{equation}
are phenomenological electron intraband spin-conserving and spin-flip relaxation rates.
In a recent pump-probe experiment on a CdSe/ZnSe quantum dot \cite{Sotier2009}, the two relaxation
rates have been estimated to be of the order of $\Gamma_{\rm s}\sim 0.01 \, {\rm ps}^{-1}$ and 
$\Gamma_{\rm c}\sim 0.1 \, {\rm ps}^{-1}$ respectively,
corresponding to two well separated time scales.

Hole spin relaxation has been found to be much slower, of the order of 
$\tau^h_s=1/\Gamma^h_{\s\bar{\s}}\sim 20$ ns 
\cite{Flissikowski2003} in both CdSe and InAs quantum dots and can be safely ignored here.
We will only consider hole charge relaxation $\Gamma_{\rm h}=\Gamma^h_{\s\s}$.


\section{Equations of motion}

To compare with pump-probe experiments, 
we calculate the differential transmission signal $\Delta T/T$
\begin{equation}
\frac{\Delta T}{T}(\tau,\omega) = \frac{T_{\rm on}-T_{\rm off}}{T_{\rm off}}
\propto {\rm Im}[P^{(3)}(\omega)] ,
\end{equation}
where
$T_{\rm on}$ ($T_{\rm off}$) is the probe pulse transmission coefficient
when the pump pulse is on (off), and $P^{(3)}$ is the induced polarization
in frequency space in third order in the optical field.

The polarization is connected with the off-diagonal density matrix elements
\begin{equation}
P= d \sum_{n\s} P_{n\s} ,
\end{equation}
where
\begin{equation}
P_{n\s}=\la \op{P}_{n\s}\ra = \la \op{h}_{n\bar{\s}}\op{e}_{n\s}\ra \equiv {\rm Tr}[ \op{h}_{n\bar{\s}}\op{e}_{n\s} \rho]
\label{Pn}
\end{equation}
describes the interband excitation of an $e$-$h$ pair in level $n$ with spins $\s$ and 
$\bar{\s}$ respectively.
Here, we have introduced the average $\la\cdots \ra \equiv  {\rm Tr}[ \cdots \rho]$.
Using \eq{ME} with only two QD levels (s and p) per band and factorizing all
four-operator expectation values within the Hartree-Fock approximation, the
polarization dynamics is described by
\begin{eqnarray}
i\dot P_{n\s}
&=& 
(E^e_{n\s}+E^h_{n\s}-V^{eh}_{nn}-i\gamma_P) P_{n\s}
\label{PneqG}\\
&&- d E(t)(1-N^e_{n\s}-N^h_{n\s})
\nonumber\\
&&+
P_{n\s}\sideset{}{'}{\sum}_{m\s'}
U_{nm} (N^e_{m\s'} + N^h_{m\bar{\s'}})
\nonumber\\
&&-
i\frac{P_{n\s}}{2}\sum_{\s'}\Gamma^e_{\s\s'}
\left[\delta_{ns}N^e_{p\s'}+\delta_{np}(1-N^e_{s\s'})\right]
\nonumber\\
&&-
i\frac{P_{n\s}}{2}\Gamma_h
\left[\delta_{ns}N^h_{p\bar{\s}}+\delta_{np}(1-N^h_{s\bar{\s}})\right]
\nonumber
\end{eqnarray}
where
\begin{equation}
N^e_{n\s} = \la \opd{e}_{n\s}\op{e}_{n\s} \ra,\qquad
N^h_{n\s}= \la \opd{h}_{n\s}\op{h}_{n\s} \ra ,
\end{equation}
are electron and hole populations, and we have defined
$U_{nm}=V^{ee}_{nm}-V^{eh}_{nm}=V^{hh}_{nm}-V^{eh}_{nm}$. The primed summation
runs over all states $\{m\s'\}\ne\{n\s\}$, and 
polarization dephasing is described with a phenomenological 
dephasing rate $\gamma_P$. 

The first three terms of \eq{PneqG} correspond to the semiconductor 
Bloch equations \cite{HaugKoch2009}. 
The second term is the standard phase space filling term due to Pauli blocking,
while the third term describes the renormalization of single particle energies
due to Coulomb interactions.
The last two terms describe a population-dependent dephasing of polarization due
to electron and hole relaxation.

The dynamics of electron and hole populations is described by similar equations of motion,
\begin{multline}
i\dot N^r_{n\s}
=
-i\gamma_N N^r_{n\s}- d E(t) P_{n\s}^\ast+ d^\ast E^\ast(t) P_{n\s}
\\
+i\delta_{re}
\sum_{\s_1\s_2}\Gamma^e_{\s_1\s_2}N^e_{p\s_1}(1-N^e_{s\s_2})
(\delta_{ns}\delta_{\s_2\s}-\delta_{np}\delta_{\s_1\s}),
\\
+i\delta_{rh}
\Gamma_{h}N^h_{p\s}(1-N^h_{s\s})
(\delta_{ns}-\delta_{np}),
\label{NeqG}
\end{multline}
with $r=e,h$ and $\gamma_N$ the population relaxation rate. 
Again, the last two lines in \eq{NeqG} describe the effect of
intraband $p\rightarrow s$ shell relaxation.

Since in pump-probe experiments the measurable quantities are at least
third order in the optical field, the above equations may be expanded in terms of
increasing order in $E(t)$, i.e. $P_{n\s}=P^{(1)}_{n\s}+P^{(3)}_{n\s}+O(E^5)$
and $N^r_{n\s}=N^{r(0)}_{n\s}+N^{r(2)}_{n\s}+O(E^4)$. Note that $N^{r(0)}_{n\s}$ is essentially
the ground state population, which vanishes for undoped systems. In our case, assuming
that the ground state electron lies in the lowest QD level, 
$N^{r(0)}_{n\s} = \nu^{r}_{n\s}=\delta_{re}\delta_{ns}\nu^e_{s\s}$ 
where $\nu^e_{s\s}$ is the s-shell filling factor.

In this manner we obtain a closed set of equations up to third order in the optical field, 
which are written explicitly in Appendix A. In the next section we will discuss their
analytical and numerical solutions and calculate the differential transmission signal.


\section{Results and Discussion}

\subsection{Analytical Solutions}

\begin{figure}[t]
\includegraphics[width=0.7\linewidth]{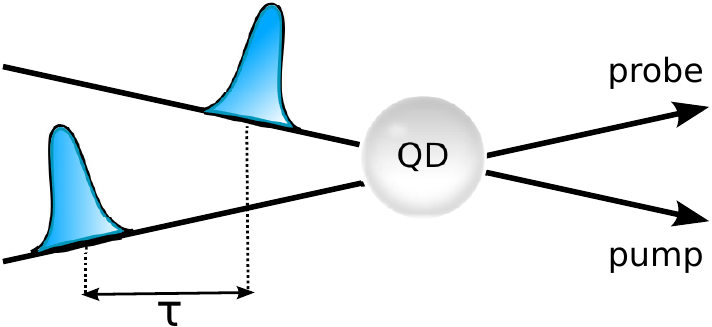}
\caption{
Schematic representation of a typical pump-probe setup. The system is
photoexcited by a strong pump pulse followed by a weaker probe pulse after 
time delay $\tau$. The signal emitted in the direction of the probe pulse is
measured as a function of $\tau$.
}
\label{pulses}
\end{figure}

The equations derived in the previous section can now be solved numerically
for any exciting laser field $E(t)$. In the special case of ultrashort pump and probe
pulses that can be described by delta functions, Eqs.(\ref{P1})-(\ref{P3}) 
can be solved analytically. Even though in
this case all QD levels can be excited (which is not the case in the
experiment), 
analytical expressions provide useful insight for the dynamics 
and we will discuss them briefly in this section.

We assume an optical field that consists of two laser pulses propagating
with time delay $\tau$ with respect to each other, i.e. it has the 
following form (at the QD):
\begin{equation}
E(t) = E_{\rm probe}(t) 
+ E_{\rm pump}(t+\tau) 
\label{Efield}
\end{equation}
where
$E_{i}(t)=E_{0}^{i}\delta(t)$, $i={\rm pump},{\rm probe}$, and 
$E_{0}^{\rm probe}$ ($E_{0}^{\rm pump}$) is the amplitude of the probe (pump) pulse
that 
arrives at the system at time $t=0$ ($t=-\tau$)
(\fig{pulses}).

Using \eq{Efield} in the equations of motion (for details see Appendix A), we obtain
the interband polarization in first order in the optical field,
\begin{multline}
P_{n\s}^{(1)}(t) = 
i d (1-\nu_{n\s}^e)
\left[E_0^{\rm probe}
 e^{-iE_{n\s}t}e^{-\gamma_{n\s}t}\theta(t)
\right.\\
\left.
+E_0^{\rm pump}
e^{-iE_{n\s}(t+\tau)}e^{-\gamma_{n\s}(t+\tau)}\theta(t+\tau)
\right]
\end{multline}
which consists of two parts due to the two pulses in the optical field.

For quantities that are second or third order in the optical field,
we will only retain terms that are up to first order in the probe pulse,
assuming that it is much weaker than the pump
($E_0^{\rm probe}\ll E_0^{\rm pump}$). In this case, 
the solution for the hole population, \eq{Nh}, has the form
\begin{multline}
N_{n\s}^h(t)=
| d|^2(1-\nu^e_{n\s}) E_0^{\rm pump}\\
\left\{
E_0^{\rm pump}
e^{-\gamma^{Nh}_n(t+\tau)}\theta(t+\tau)
+E_0^{\rm probe} e^{i E_{n\s}\tau}e^{-\gamma_{n\s}|\tau|}
\right.\\ 
\left.
\times\left[\theta(\tau) e^{-\gamma^{Nh}_n t}\theta(t)
+\theta(-\tau)e^{-\gamma^{Nh}_n (t+\tau)}\theta(t+\tau)
\right]\right\}
\label{solNhn}
\end{multline}
which describes the creation of hole population in the
$n$-th shell either from the pump pulse only, or from
both the pump and probe pulses. Here we defined
$\gamma^{Nh}_n=\gamma_N+(\delta_{np}-\delta_{ns})\Gamma_{h}$.

For the electronic populations we obtain similar expressions,
but $\gamma_N$ is replaced by a level-dependent relaxation rate
$\gamma_{n\s}^{Ne}=
\gamma_N+\delta_{np}(\nu^e_{n\bar{\s}}\Gamma_{\rm c}+\nu^e_{n\s}\Gamma_{\rm s})$
and there are additional terms
of the form
\[
\delta_{ns}\sum_{\s'}
\left(e^{-\gamma_{s\s}^{Ne} t}-e^{-\gamma_{p\s'}^{Ne} t}\right)
\]
that describe the rise of the $s$-shell electron population due to interlevel
relaxation. 
These terms also appear in the solution for the third order terms $P_{n\s}^{(3)}$,
and lead for a 
spin-dependent increase of the differential transmission signal as a function
of the time delay.
The exact expressions for $N^{e(2)}_{n\s}$ and $P_{n\s}^{(3)}$ are 
included in Appendix B.


\subsection{Zero magnetic field}

\begin{figure}[t]
\includegraphics[width=\linewidth]{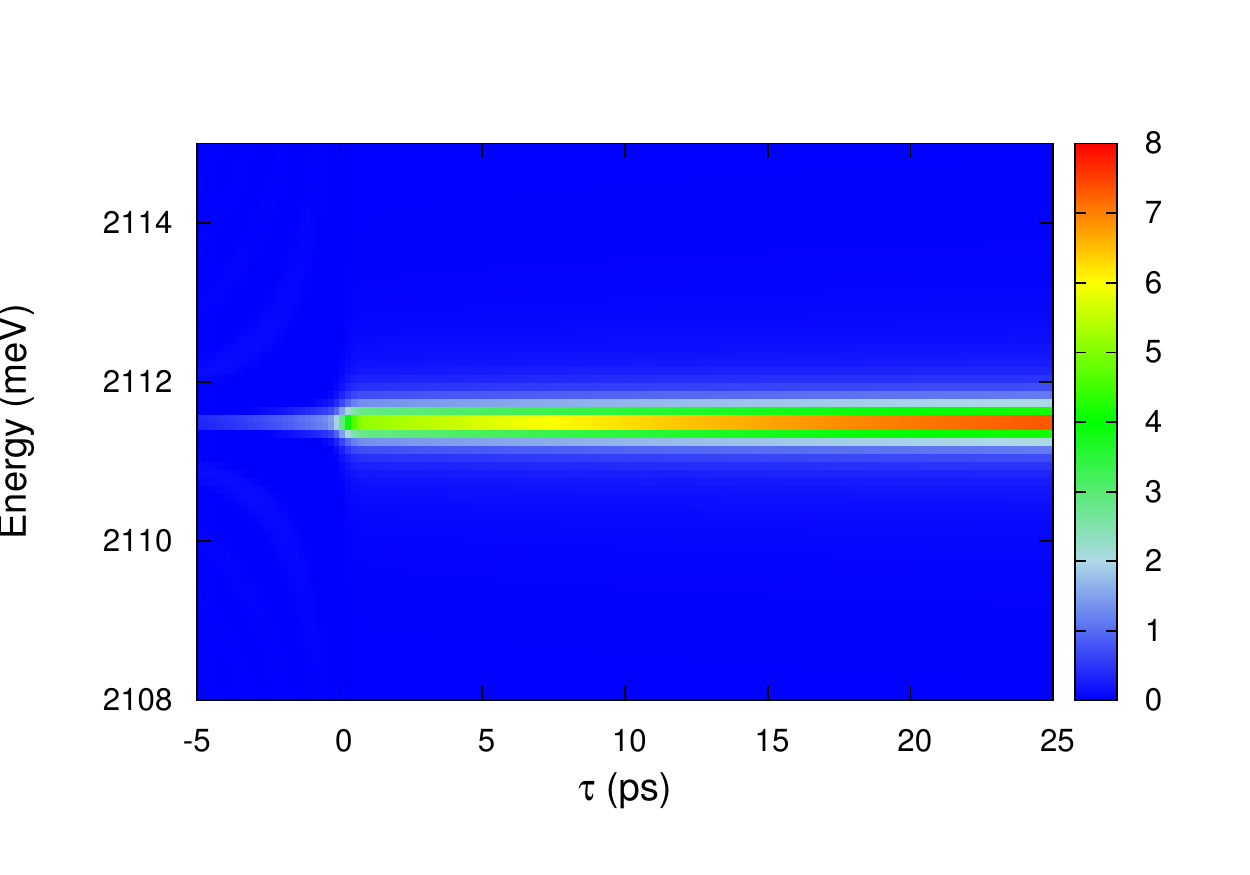}
\caption{
Differential transmission signal $\Delta T/T$ (in arbitrary units) 
as a function of time delay $\tau$
and probe pulse energy $\hbar\omega$ for an unpolarized ground state
electron, $\nu^e_{s\uparrow}=\nu^e_{s\downarrow}=0.5$,
and linearly polarized, Gaussian pulses
with duration $T_{\rm pump}=700$ fs, $T_{\rm probe}=180$ fs.
For this plot, we have used the parameters $E_s=2110$ meV, $E_p=2210$ meV, 
$\gamma_P=5$ ps, $\Gamma_{\rm h}=0.1$ ps, $\Gamma_{\rm c}=15$ ps, $\Gamma_{\rm s}=170$ ps, 
$\gamma_N=480$ ps, and $U_{nm}=0$.
}
\label{Fig4}
\end{figure}

In this section, we discuss the results from our numerical calculations of
the differential transmission signal for Gaussian pulses 
similar to the experiment of Ref.~\onlinecite{Sotier2009}.
\fig{Fig4} shows the imaginary part of the nonlinear polarization $P^{(3)}(\omega)$ 
for the case of an unpolarized ground state electron as a function of the time delay 
$\tau$ between the pump and probe pulse and the probe pulse energy. There is a single peak
at the $s$-shell trion energy $E_s$ that increases with time delay for tens of
picoseconds, in agreement with the experimental findings. This slow increase
of the signal is a signature of intraband relaxation from the $p$ to the $s$ shell,
for which spin-conserving and spin-flipping mechanisms contribute, since the
optical pulses are linearly polarized and the ground state electron unpolarized.

\begin{figure}
\includegraphics[width=\linewidth]{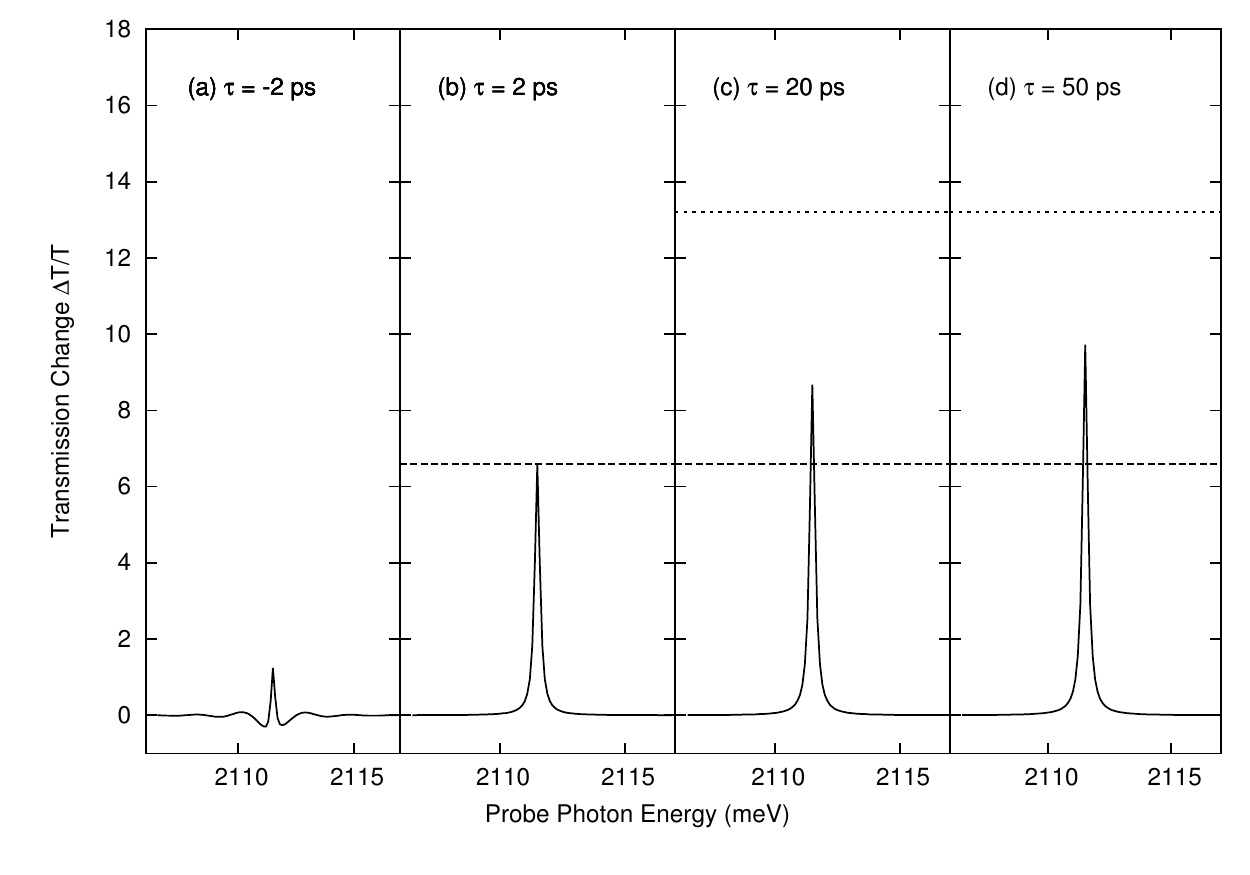}
\caption{
Differential transmission signal as a function of probe pulse energy for different time delays $\tau$.
All parameters are as in \fig{Fig4}. The dashed line marks the signal right after excitation by
the pump pulse, that corresponds to bleaching due to hole interlevel relaxation. The dotted line
marks the expected signal for full electronic relaxation.
}
\label{Fig5}
\end{figure}

\begin{figure}
\includegraphics[width=\linewidth]{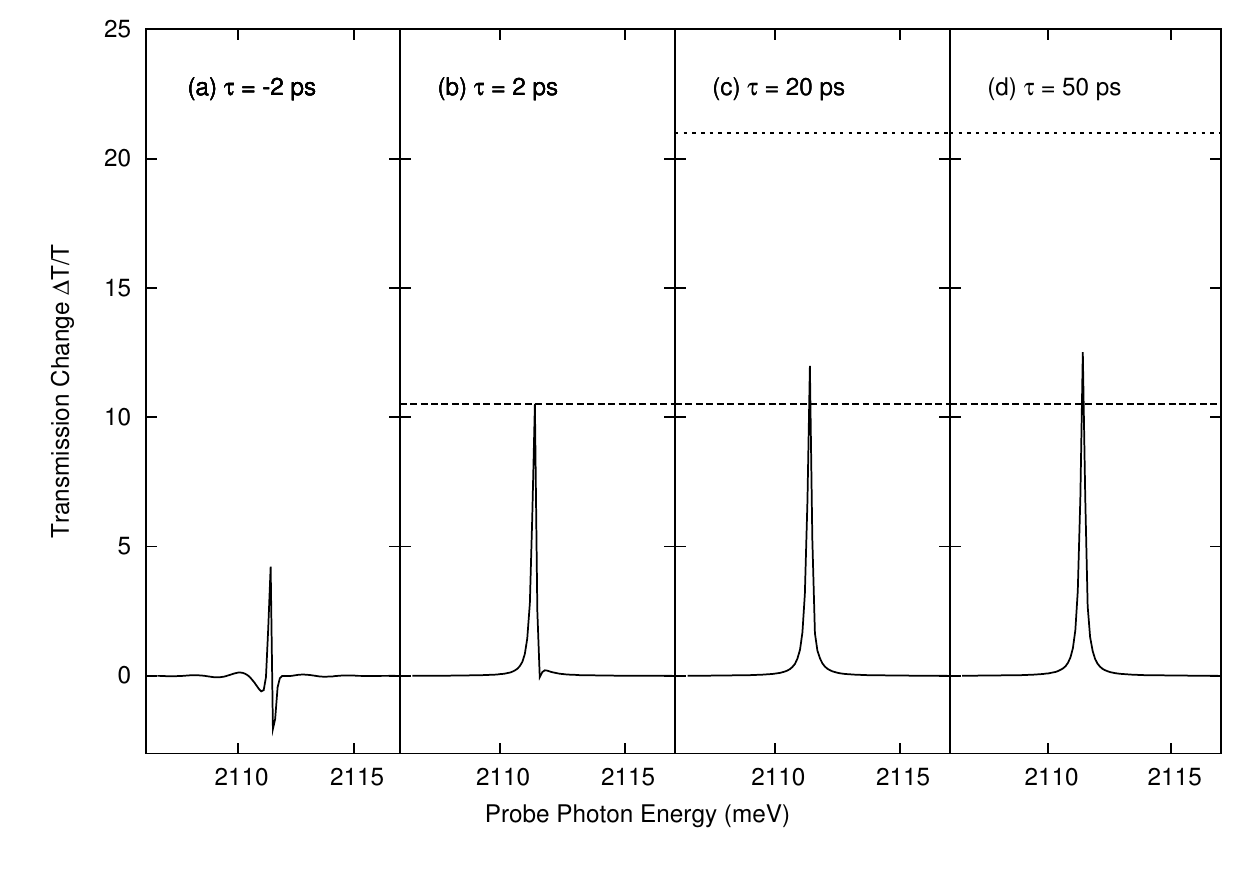}
\caption{
Same as \fig{Fig5} but including Coulomb interactions ($U_{nm}=0.1$ meV).
}
\label{Fig6}
\end{figure}

\begin{figure}
\includegraphics[width=0.8\linewidth]{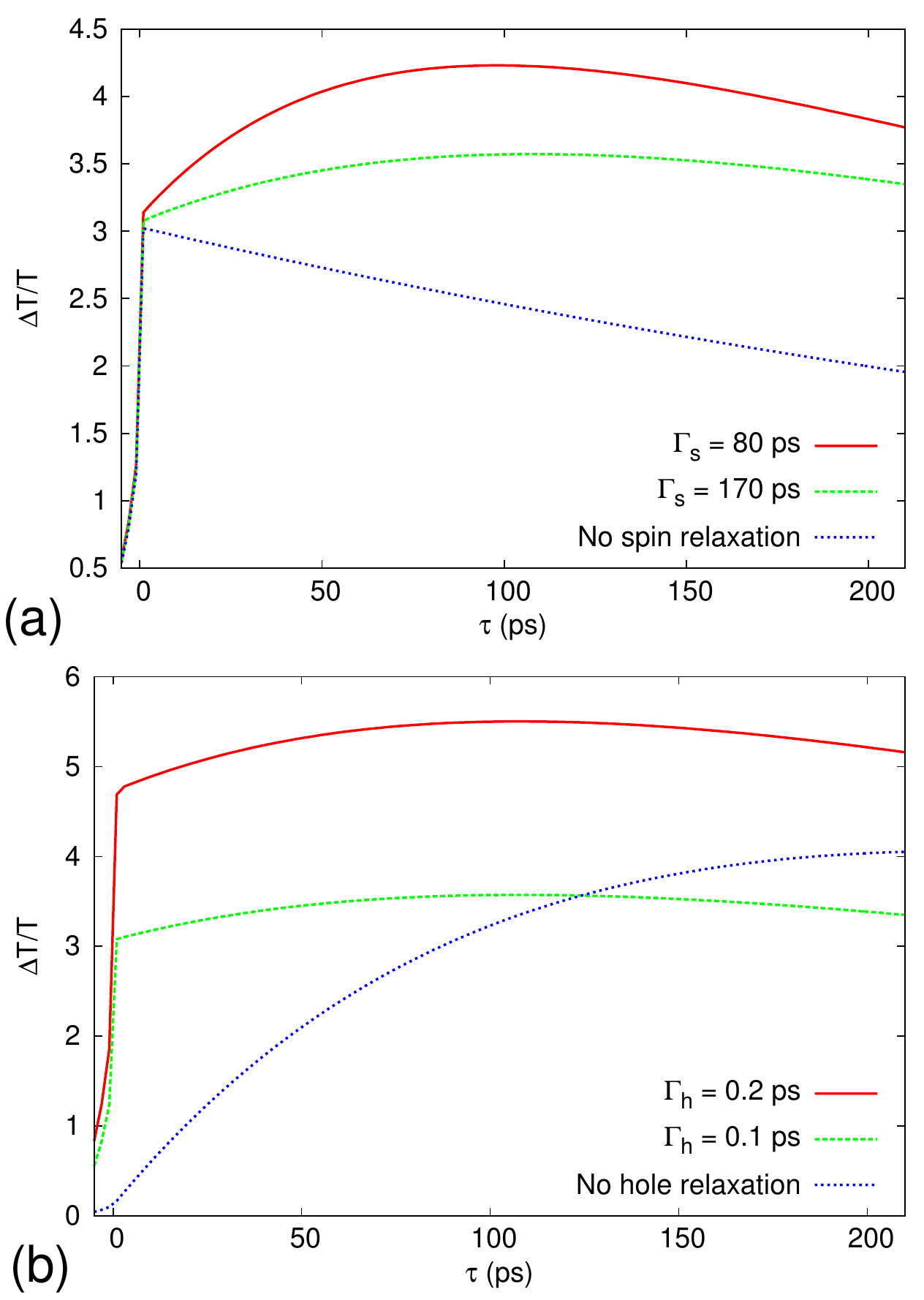}
\caption{
Differential transmission signal as a function of time delay $\tau$,
(a) with and without the electron spin relaxation and
(b) with and without hole relaxation 
for a $\s=\downarrow$ 
ground state electron and $\s_+$ polarized pump pulse.
All other parameters are as in \fig{Fig4}.
}
\label{Fig7}
\end{figure}

A more detailed description of the dynamics is shown in \fig{Fig5} which depicts snapshots of the signal for specific time delays. For $\tau=-2$ ps, for which the probe pulse precedes the pump, there is a small signal that arises from the interference between the two pulses and is characterized by oscillations with frequency $E_p-E_s$. 

For $\tau=2$ ps, when the probe pulse arrives right after the pump, the situation is different as the pump pulse has created an $e$-$h$ pair in the $p$-shell. The hole relaxes almost immediately to the $s$ shell in the valence band, and as the probe pulse arrives, it can either recombine with the ground state electron, or block the probe pulse absorption (bleaching), thus leading to an increase in the transmission.

For $\tau=20$ ps, the electron has relaxed to the $s$-shell only if it is in the singlet configuration. Given that only two of the triplet states are bright, this leads to an additional increase of the signal by a factor of $\sim 1.3$ in comparison to the signal at $\tau=2$ ps, that remains constant for tens of picoseconds until spin flips can take place. 

In the above results we have ignored the role of Coulomb interactions. Their contribution is shown in \fig{Fig6}, which shows the differential transmission signal for the same parameters as in \fig{Fig5}, but with the additional terms arising from Coulomb interactions. It is clear that their main effect is a shift of the fundamental trion resonance for very short timescales, but their role is diminished for larger time delays. This is in agreement with the results of Ref.~\onlinecite{Huneke2011}, where the role of Coulomb correlations has been studied.

\fig{Fig7}(a) shows the effect of electron spin relaxation on the differential
transmission $\Delta T/T$ for a $\s_+$ pump pulse and a $\s=\downarrow$ ground state electron
(in which case spin relaxation is necessary for interlevel relaxation).
For small delay times $\tau$, the slow spin-flip processes do not contribute and the
signal exhibits a sharp increase due to hole relaxation. At larger time scales,
the role of spin relaxation becomes evident by the slowly increase of the signal, the 
absolute maximum of which depends on the spin relaxation rate. As shown by the dotted
line in \fig{Fig7}(a), there is no increase of the signal in the absence of spin
relaxation.

On the other hand, hole relaxation plays an important role at early timescales.
This is shown in \fig{Fig7}(b) where the differential transmission signal is plotted
for different values of $\Gamma_h$. In the absense of hole relaxation, the signal
increases slowly due to electron spin relaxation.

 
\subsection{Finite magnetic field}

\begin{figure}
\includegraphics[width=\linewidth]{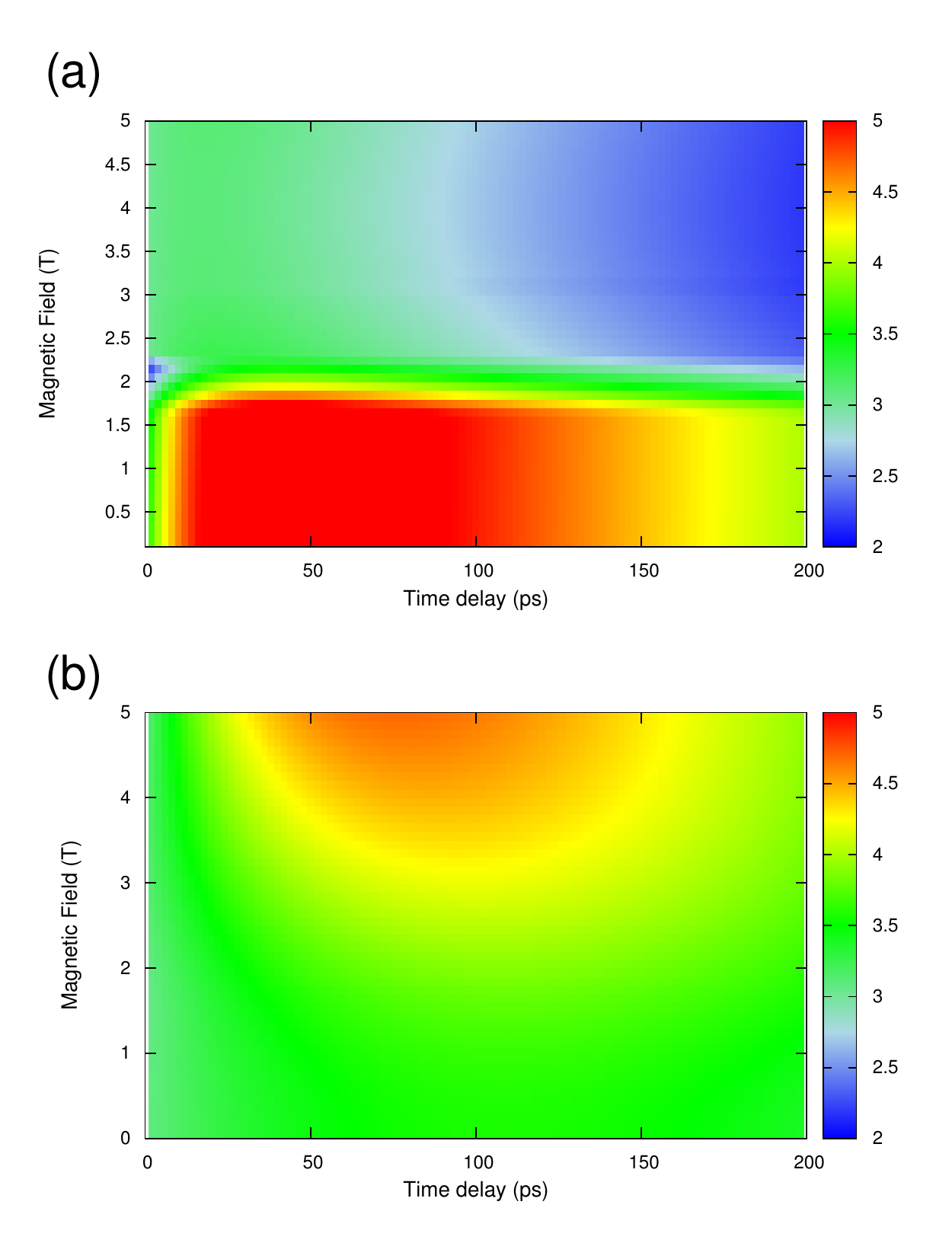}
\caption{Differential transmission
$\Delta T/T$ as a function of pump-probe delay time 
$\tau$ and magnetic field $B$ for a circularzly polarized
(a) $\s_+$ and (b) $\s_-$ pump pulse. 
We assume the temperature is low enough for the ground state electron
to be polarized. 
All other parameters are as in \fig{Fig4}.
In (a), the large spin-blockade signal for low
magnetic fields is strongly suppressed at higher
fields.
}
\label{Bfield}
\end{figure}

In the presence of an external magnetic field, 
more spin relaxing mechanisms
are allowed, thus enhancing the spin-flipping relaxation rate. 
It has been shown in Ref.~\onlinecite{Khaetskii00PRB} 
that spin relaxation in QDs is produced by a
variety of mechanisms that can be separated in two groups: 
direct spin-phonon coupling, and admixture mechanisms due to
spin-orbit coupling. In both cases though,
the finite magnetic field leads leads to a $\sim B^2$ dependence 
of the spin relaxation rate between different orbitals.
For the quantum dots considered here\cite{Sotier2009,Mahapatra2007}
and magnetic fields up to 5 T, Zeeman splitting is much smaller
($\sim\mu$eV)\cite{Hogele2005} than the interlevel spacing ($50-100$~meV) 
and its role is insignificant. Thus, the admixture of different spin states
plays a lesser role and the dominant spin flipping mechanism is the
direct spin-phonon coupling.

In \fig{Bfield}, the differential transmission signal at the $s$-shell 
resonance is shown as a function of time delay $\tau$ and magnetic
field $B$ for right and left circularly polarized pump pulses. Assuming
that the temperature is low enough for the ground state electron to be
fully polarized by the applied magnetic field, a $\sigma_-$ pulse leads
to a well defined spin blockade regime, as shown in \fig{Bfield}(b).
For low magnetic fields, the differential transmission signal is much
smaller in comparison to \fig{Bfield}(a) where spin conserving relaxation
taked place.
However, due to the $\sim B^2$ enhancement of the spin flipping rate, at
larger magnetic field spin blockade is suppressed. This in contrast with
transport experiments where the application of an external magnetic
field suppresses the singlet-triplet mixing and thus
enhances the spin blockade effect \cite{Koppens2005}.


\section{Conclusions}

We have developed a model describing the trion and population dynamics
in a photoexcited quantum dot in a pump-probe setup. We have included the
role of intersubband relaxation including spin flipping and separated its
role from the spin conserving mechanism. The long timescale of intraband 
spin relaxation leads to a signature in the differential transmission signal
that is analogous to optical spin blockade. In the
presence of an external magnetic field, the enhancement of spin-flipping
relaxation rate leads to lifting of spin blockade at shorter time
scales.
This mechanism opens new possibilites for the study of spin decoherence 
processes in semiconductor quantum dots with optical probes.


\begin{acknowledgements}
 We acknowledge useful discussions with A. Leitenstorfer and funding
 from the Konstanz Center for Applied Photonics (CAP) and from
BMBF under the program QuaHL-Rep.
\end{acknowledgements}


\appendix

\section{Equations of motion}

Here we write the set of equations, derived from Eqs. (\ref{PneqG}) and (\ref{NeqG}) 
expanded in increasing orders of the optical field. 
Keeping terms up to first order, \eq{PneqG} for the polarization becomes
\begin{equation}
i\dot{P}^{(1)}_{n\s}=(E_{n\s}-i\gamma_{n\s})P^{(1)}_{s\s}- d E(t)(1-\nu^e_{n\s}) ,
\label{P1}
\end{equation}
where $E_{n\s}=E_{n\s}^e+E_{n\s}^h-V^{eh}_{nn}+U_{ns}$ 
is the trion energy and
$\gamma_{n\s}=\gamma_P+\delta_{np}(\Gamma_{\rm c}\nu^e_{s{\bar{\s}}}
+\Gamma_{\rm s}\nu^e_{s{\s}}+\Gamma_{h})/2$
describes the trion relaxation rate, which for the $p$-shell is
enhanced by the intraband 
spin-conserving and spin-flipping relaxation terms. 

In second order in the optical field, the equations of motion 
for the electron and hole populations are
\begin{eqnarray}
i\dot{N}^{e(2)}_{s\s}&=&-i\gamma_N N^{e(2)}_{s\s}
- d E(t)P^{(1)\ast}_{s\s}+ d^\ast E^\ast(t)P^{(1)}_{s\s}
\nonumber\\
&&
+i(1-\nu^e_{s\s})\left[\Gamma_{\rm c}N^{e(2)}_{p\s}+\Gamma_{\rm s}N^{e(2)}_{p\bar{\s}}\right],
\label{Nes}
\\[2mm]
i\dot{N}^{e(2)}_{p\s}&=&-i\left[\gamma_N+(1-\nu^e_{s\s})\Gamma_{\rm c}+(1-\nu^e_{s\bar{\s}})\Gamma_{\rm s}\right] N^{e(2)}_{p\s}
\nonumber\\
&&
- d E(t)P^{(1)\ast}_{p\s}+ d^\ast E^\ast(t)P^{(1)}_{p\s},
\\[2mm]
i\dot{N}^{h(2)}_{n\s}&=&-i(\gamma_N+(\delta_{np}-\delta_{ns})\Gamma_h) N^{h(2)}_{n\s}
\nonumber\\
&&
- d E(t)P^{(1)\ast}_{n\s}+ d^\ast E^\ast(t)P^{(1)}_{n\s}.
\label{Nh}
\end{eqnarray}

Finally, for the polarization in third order, we obtain
\begin{eqnarray}
i\dot{P}^{(3)}_{n\s}&=&(E_{n\s}-i\gamma_{n\s})P^{(3)}_{n\s}
+ d E(t)\left[N^{e(2)}_{n\s}+N^{h(2)}_{n\s}\right]
\nonumber\\
&&
+P^{(1)}_{n\s}\sideset{}{'}{\sum}_{m\s'} U_{nm} (N^{e(2)}_{m\s'}+N^{h(2)}_{m\s'})
\nonumber\\
&&
+i\frac{1}{2}P_{n\s}^{(1)}\Gamma_{h}
(\delta_{np}-\delta_{ns})N^{h(2)}_{\bar{n}\bar{\s}}
\nonumber\\
&&
+i\frac{1}{2}P_{n\s}^{(1)}\sum_{\s'}\Gamma^e_{\s\s'}
\left(\delta_{np}N^{e(2)}_{s\s'}-\delta_{ns}N^{e(2)}_{p\s'}\right)
\label{P3}
\end{eqnarray}
The last term in the above equation describes contributions
from interlevel relaxation of electronic populations, 
which as discussed in section IV, leads to spin-dependent 
signatures in the differential transmission signal.


\section{Analytical Solutions}

The solution for the electronic populations has the form
\begin{multline}
N_{n\s}^e(t)
=
| d|^2 (E_0^{\rm pump})^2(1-\nu_{n\s}^e)\theta(t+\tau)
\left\{
e^{-\gamma_{n\s}^{Ne}(t+\tau)}
\right.
\\
+\delta_{ns}\sum_{\s'} a_{\s\s'}
\left.\left[e^{-\gamma_{s\s}^{Ne}(t+\tau)}-e^{-\gamma_{p\s'}^{Ne}(t+\tau)}\right]
\right\}\\
+2| d|^2 E_0^{\rm pump}E_0^{\rm probe}
(1-\nu_{n\s}^e)
\Big\{\cos(E_{n\s}\tau) e^{-\gamma_{n\s}|\tau|}\\
\times\left[e^{-\gamma_{n\s}^{Ne} t}\theta(\tau)\theta(t)
+e^{-\gamma_{n\s}^{Ne} (t+\tau)}\theta(-\tau)\theta(t+\tau)\right]\\
+\delta_{ns}
\sum_{\s'} a_{\s\s'}
\cos(E_{p\s'}\tau)e^{-\gamma_{p\s'}|\tau|}\\
\left[
\theta(\tau)\theta(t)(e^{-\gamma_{s\s}^{Ne} t}-e^{-\gamma_{p\s'}^{Ne} t})
\right.\\
\left.
+\theta(-\tau)\theta(t+\tau)(e^{-\gamma_{s\s}^{Ne} (t+\tau)}-e^{-\gamma_{p\s'}^{Ne} (t+\tau)})
\right]\Big\}
\label{solNen}
\end{multline}
where
$\gamma_{n\s}^{Ne}=
\gamma_N+\delta_{np}(\nu^e_{n\bar{\s}}\Gamma_{\rm c}+\nu^e_{n\s}\Gamma_{\rm s})$. 
and
$a_{\s\s'}=\Gamma^{\s\s'}_{\rm sp}/(\nu_{s\bar{\s}'}\Gamma_c+\nu_{s\s'}\Gamma_s)$.
Comparing the above expression with the solution for the hole populations,
\eq{solNhn}, there are additional terms ($\propto \delta_{ns}$)
that describe the creation of electronic population in the $s$-shell due 
to intraband relaxation.

The solution for the third order terms $P_{n\s}^{(3)}$,
which contribute to the differential transmission signal, is
given by (for $U_{nm}=0$)
\begin{multline}
P_{n\s}^{(3)}(\omega)
=(1-\nu^e_{n\s})
\frac{ d E_{0}^{\rm probe}(E_0^{\rm pump})^2}{\omega-E_{n\s}+i\gamma_{n\s}}\\
\left[
| d|^2 (e^{-\gamma^{Nh}_n\tau}+e^{-\gamma^{Ne}_{n\s}\tau})
\right.\\
+\delta_{ns}\sum_{\s'} a_{\s\s'}
\left(e^{-\gamma^{Ne}_{s\s}\tau} - e^{-\gamma^{Ne}_{p\s'}\tau} \right)\\
+\frac{1}{2}i(\delta_{ns}-\delta_{np})\sum_{\s'}
\frac{\Gamma^{\s\s'}_{\rm sp}| d|^2}{\omega-E_{n\s}+i(\gamma_{n\s}+\gamma^{Ne}_{p\s'})}\\
\left.
\times\left(e^{-\gamma^{Ne}_{p\s'}\tau}+e^{-i(E_{n\s}-E_{p\s'})\tau}e^{-(\gamma_{n\s}+\gamma^{Ne}_{p\s'})\tau}\right)
\right]
\end{multline}
for $\tau>0$ and
\begin{multline}
P_{n\s}^{(3)}(\omega)
=(1-\nu^e_{n\s})
\frac{ d E_{0}^{\rm probe}(E_0^{\rm pump})^2}{\omega-E_{n\s}+i\gamma_{n\s}}\\
\left[
| d|^2 e^{-i(\omega-E_{n\s})\tau} e^{\gamma_{n\s}\tau})
+\frac{1}{2}i(\delta_{ns}-\delta_{np})
\right.\\
\times\sum_{\s'}\frac{\Gamma^{\s\s'}_{\rm sp}|d|^2 e^{-i(\omega-E_{p\s'})\tau}}
{\omega-E_{n\s}+i(\gamma_{n\s}+\gamma^{Ne}_{p\s'})}
\left(e^{\gamma_{n\s}\tau}+e^{\gamma_{p\s'}\tau}\right)
\bigg]
\end{multline}
for $\tau<0$.
Given that $\gamma_{n\s}\gg\gamma_{n\s}^{Ne}$,
it is clear from the above solution that the differential transmission signal
decays fast for $\tau<0$, while for $\tau>0$ it is dominated 
by the $\sim e^{-\gamma_{n\s}^{Ne} \tau}$ term at long timescales, leading to 
spin-dependent decay.


\bibliography{OpticalSpinBlockade}

\end{document}